
\newcount\machine
\global\machine=2   
\magnification1200
\tolerance=10000
\hsize=17truecm\vsize=23truecm
\parindent=40pt
\mathsurround=0pt
\baselineskip=16pt plus 2pt minus 1pt
\def\showintremarks{y}  
\long\def\intremark#1{\if\showintremarks y
                \removelastskip\vskip.1in\hrule\penalty500\vskip.05in\noindent
                {\bf Internal Remark}\hfill\break\penalty500
                 #1
                 \vskip.05in\penalty500\hrule\vskip.1in\penalty-500
                 \fi}
%
%
%
%
%
%
\newdimen\rulesz \newdimen\bordersz \newdimen\boxshift
\rulesz=1pt \bordersz=2pt
\def\shbox#1{\setbox111=\hbox{#1\hss}%
\boxshift=\dp111\advance\boxshift by \rulesz\advance\boxshift by \bordersz%
\lower\boxshift\vbox{\hrule height\rulesz%
\hbox{\vrule width\rulesz\kern\bordersz%
\vbox{\kern\bordersz\copy111\kern\bordersz}\kern\bordersz%
\vrule width\rulesz}\hrule height\rulesz}}%
%
%
\def\today{\ifcase\month\or January\or February\or March\or April\or
     May\or June\or July\or August\or September\or October\or November\or
     December\fi\space\number\day, \number\year}
\def\header#1{\rm\nopagenumbers \hfil\underbar{#1} \hfil\today\bigskip
     \headline={\rm\ifnum\pageno>1 #1\hfil\today\hfil Page
\folio\else\hfil\fi}}
\def\dst{\displaystyle}

\def\sst{\scriptstyle}

\def\tst{\textstyle}

\def\frac#1#2{\dst {#1\over#2}}     
\def\sfrac#1#2{{\tst{#1\over#2}}}   
\def\pmb#1{\setbox0=\hbox{#1}       
     \kern-.025em\copy0\kern-\wd0
     \kern.05em\copy0\kern-\wd0
     \kern-.025em\box0}             
%
%
\def\al{\alpha}
\def\be{\beta}
\def\ga{\gamma}
\def\de{\delta}

\def\et{\eta}

\def\la{\lambda}

\def\si{\sigma}

\def\Ga{\Gamma}
\def\De{\Delta}

\def\La{\Lambda}

%
%
\def\0{{\bf 0}}

\def\k{{\bf k}}

\def\m{{\bf m}}

\def\q{{\bf q}}
\def\x{{\bf x}}

\def\p{{\bf p}}

\def\bpsi{{\bf \Psi}}

\def\cE{{\cal E}}

\def\cS{{\cal S}}

%
%

\def\FF{\hbox to 8.33887pt{\rm I\hskip-1.8pt F}}
\def\NN{\hbox to 9.3111pt{\rm I\hskip-1.8pt N}}
\def\PP{\hbox to 8.61664pt{\rm I\hskip-1.8pt P}}
\def\QQ{\rlap {\raise 0.4ex \hbox{$\scriptscriptstyle |$}}
{\hskip -4.5pt Q}}
\def\RR{\hbox to 9.1722pt{\rm I\hskip-1.8pt R}}

%
%
%
\font\tenfrak=eufm10\font\sevenfrak=eufm7\font\fivefrak=eufb5
\newfam\frakfam
     \textfont\frakfam=\tenfrak
     \scriptfont\frakfam=\sevenfrak
     \scriptscriptfont\frakfam=\fivefrak

\def\bbbr{{\rm I\!R}}

\font \tensans                = cmss10
\font \fivesans               = cmss10 at 5pt

\font \sevensans              = cmss10 at 7pt

\font \twelvesans             = cmss12
\font \smallescriptscriptfont = cmr5
\font \smallescriptfont       = cmr5 at 7pt
\font \smalletextfont         = cmr5 at 10pt
\newfam\sansfam
\textfont\sansfam=\tensans\scriptfont\sansfam=\sevensans
                   \scriptscriptfont\sansfam=\fivesans
\def\sans{\fam\sansfam\tensans}
\def\bbbone{{\mathchoice {\rm 1\mskip-4mu l} {\rm 1\mskip-4mu l}    
{\rm 1\mskip-4.5mu l} {\rm 1\mskip-5mu l}}}
\def\bbbc{{\mathchoice {\setbox0=\hbox{$\displaystyle\rm C$}\hbox{\hbox 
to0pt{\kern0.4\wd0\vrule height0.9\ht0\hss}\box0}}
{\setbox0=\hbox{$\textstyle\rm C$}\hbox{\hbox
to0pt{\kern0.4\wd0\vrule height0.9\ht0\hss}\box0}}
{\setbox0=\hbox{$\scriptstyle\rm C$}\hbox{\hbox
to0pt{\kern0.4\wd0\vrule height0.9\ht0\hss}\box0}}
{\setbox0=\hbox{$\scriptscriptstyle\rm C$}\hbox{\hbox
to0pt{\kern0.4\wd0\vrule height0.9\ht0\hss}\box0}}}}
\def\bbbe{{\mathchoice {\setbox0=\hbox{\smalletextfont e}\hbox{\raise   
0.1\ht0\hbox to0pt{\kern0.4\wd0\vrule width0.3pt
height0.7\ht0\hss}\box0}}
{\setbox0=\hbox{\smalletextfont e}\hbox{\raise
0.1\ht0\hbox to0pt{\kern0.4\wd0\vrule width0.3pt
height0.7\ht0\hss}\box0}}
{\setbox0=\hbox{\smallescriptfont e}\hbox{\raise
0.1\ht0\hbox to0pt{\kern0.5\wd0\vrule width0.2pt
height0.7\ht0\hss}\box0}}
{\setbox0=\hbox{\smallescriptscriptfont e}\hbox{\raise
0.1\ht0\hbox to0pt{\kern0.4\wd0\vrule width0.2pt
height0.7\ht0\hss}\box0}}}}
\def\bbbq{{\mathchoice {\setbox0=\hbox{$\displaystyle\rm               
Q$}\hbox{\raise
0.15\ht0\hbox to0pt{\kern0.4\wd0\vrule height0.8\ht0\hss}\box0}}
{\setbox0=\hbox{$\textstyle\rm Q$}\hbox{\raise
0.15\ht0\hbox to0pt{\kern0.4\wd0\vrule height0.8\ht0\hss}\box0}}
{\setbox0=\hbox{$\scriptstyle\rm Q$}\hbox{\raise
0.15\ht0\hbox to0pt{\kern0.4\wd0\vrule height0.7\ht0\hss}\box0}}
{\setbox0=\hbox{$\scriptscriptstyle\rm Q$}\hbox{\raise
0.15\ht0\hbox to0pt{\kern0.4\wd0\vrule height0.7\ht0\hss}\box0}}}}

\def\bbbs{{\mathchoice                                               
{\setbox0=\hbox{$\displaystyle     \rm S$}\hbox{\raise0.5\ht0\hbox
to0pt{\kern0.35\wd0\vrule height0.45\ht0\hss}\hbox
to0pt{\kern0.55\wd0\vrule height0.5\ht0\hss}\box0}}
{\setbox0=\hbox{$\textstyle        \rm S$}\hbox{\raise0.5\ht0\hbox
to0pt{\kern0.35\wd0\vrule height0.45\ht0\hss}\hbox
to0pt{\kern0.55\wd0\vrule height0.5\ht0\hss}\box0}}
{\setbox0=\hbox{$\scriptstyle      \rm S$}\hbox{\raise0.5\ht0\hbox
to0pt{\kern0.35\wd0\vrule height0.45\ht0\hss}\raise0.05\ht0\hbox
to0pt{\kern0.5\wd0\vrule height0.45\ht0\hss}\box0}}
{\setbox0=\hbox{$\scriptscriptstyle\rm S$}\hbox{\raise0.5\ht0\hbox
to0pt{\kern0.4\wd0\vrule height0.45\ht0\hss}\raise0.05\ht0\hbox
to0pt{\kern0.55\wd0\vrule height0.45\ht0\hss}\box0}}}}
\def\bbbz{{\mathchoice {\hbox{$\sans\textstyle Z\kern-0.4em Z$}}       
{\hbox{$\sans\textstyle Z\kern-0.4em Z$}}
{\hbox{$\sans\scriptstyle Z\kern-0.3em Z$}}
{\hbox{$\sans\scriptscriptstyle Z\kern-0.2em Z$}}}}
%
%
\def\const{{\rm const}\,}

\def\tr{{\rm tr\, }}

\def\cartprod{\mathop{\lower2pt\hbox{{\twelvesans X}}}}

\def\bbar#1{\setbox0=\hbox{$#1$}%
     \copy0\kern-\wd0
     \raise.0433em\box0}
\def\optbar#1{\vbox{\ialign{##\crcr\hfil${\scriptscriptstyle(}\mkern -1mu
         \vrule height 1.2pt width 3pt depth -.8pt
         {\scriptscriptstyle)}$\hfil\crcr
       \noalign{\kern-1pt\nointerlineskip}$\hfil\displaystyle{#1}\hfil$\crcr}}}
\def\<{\left<}
\def\>{\right>}
\def\dbar{{\mkern6mu\mathchar'26\mkern-12mud}}
\def\sbar#1{\vbox{\ialign{##\crcr
         \noalign{\kern1pt\nointerlineskip}
         \hfil$ \mkern -1mu\vrule height 1.2pt width 3pt depth -.9pt$\hfil\crcr
         \noalign{\kern0pt\nointerlineskip}
                $\hfil{\sst #1}\hfil$\crcr}}}
\def\smprod{\mathop{\textstyle\prod}}
\def\smsum{\mathop{\textstyle\sum}}

\ifnum\machine=1\font\tenmsa=msxm10\font\fivemsa=msxm5\fi
\ifnum\machine=2\font\tenmsa=msam10\font\fivemsa=msam5\fi
\ifnum\machine=1\font\tenmsb=msym10\font\sevenmsb=msym7\font\fivemsb=msym5\fi
\ifnum\machine=2\font\tenmsb=msbm10\font\sevenmsb=msbm7\font\fivemsb=msbm5\fi
      \newfam\msafam
     \textfont\msafam=\tenmsa
     \scriptfont\msafam=\fivemsa   
     \scriptscriptfont\msafam=\fivemsa
\def\msafamno{\ifcase \msafam 0\or 1\or 2\or 3\or 4\or 5\or 6\or 7\or 8\or
 9\or A\or B\or C\or D\or E\or F\fi}
\newfam\msbfam
     \textfont\msbfam=\tenmsb
     \scriptfont\msbfam=\sevenmsb
     \scriptscriptfont\msbfam=\fivemsb
\def\msbfamno{\ifcase \msbfam 0\or 1\or 2\or 3\or 4\or 5\or 6\or 7\or 8\or
 9\or A\or B\or C\or D\or E\or F\fi}

\mathchardef\lle="3\msafamno36
\mathchardef\gge="3\msafamno3E
\mathchardef\smsetminus="2\msbfamno72

\mathchardef\evall="2\msafamno16
%
%
\font \tafontt                = cmbx10 scaled\magstep2
\font \tbfontss               = cmbx5  scaled\magstep1
\font \tbfonts                = cmbx7  scaled\magstep1
\font \tbfontt                = cmbx10 scaled\magstep1
\font \tasys                  = cmex10 scaled\magstep1
\font \tamsss                 = cmmib10 scaled 833
\font \tamss                  = cmmib10
\font \tams                   = cmmib10 scaled\magstep1
\font \tbst                   = cmsy10 scaled\magstep1
\font \tbsss                  = cmsy5  scaled\magstep1
\font \tbss                   = cmsy7  scaled\magstep1

\font \subchfont              = cmbx12 at 12pt

\def\titleb#1{\removelastskip\vskip.3truein\noindent{\tbfontt #1
}\vskip.25truein\nobreak}

\def\section#1#2{%
      \vskip25pt plus 4pt minus4pt
     \bgroup
 \textfont0=\tbfontt \scriptfont0=\tbfonts \scriptscriptfont0=\tbfontss
 \textfont1=\tams \scriptfont1=\tamss \scriptscriptfont1=\tamsss
 \textfont2=\tbst \scriptfont2=\tbss \scriptscriptfont2=\tbsss
 \textfont3=\tasys \scriptfont3=\tenex \scriptscriptfont3=\tenex
     \baselineskip=16pt
     \lineskip=0pt
     \rightskip 0pt plus 6em
     \pretolerance=10000
     \tbfontt
     \setbox0=\vbox{\vskip25pt
     \noindent
     \if!#1!\ignorespaces#2
     \else\ignorespaces#1\unskip\enspace\ignorespaces#2\fi
     \vskip15pt}%
     \dimen0=\pagetotal
     \ifdim\dimen0<\pagegoal
     \dimen0=\ht0\advance\dimen0 by\dp0\advance\dimen0 by
     4\normalbaselineskip
     \advance\dimen0 by\pagetotal
     \advance\dimen0 by-\pageshrink
     \ifdim\dimen0>\pagegoal\vfill\eject\fi\fi
     \noindent
     \if!#1!\ignorespaces#2
     \else\ignorespaces#1\unskip\enspace\ignorespaces#2\fi
     \egroup
     \vskip12.5pt plus4pt minus4pt
     \ignorespaces}
%
%
\def\newenvironment#1#2#3#4{\long\def#1##1##2{\removelastskip\penalty-100
\vskip\baselineskip\noindent{#3#2\if!##1!.\else\unskip\ \ignorespaces
##1\unskip\fi\ }{#4\ignorespaces##2\vskip\baselineskip}}}
\newenvironment\lemma{Lemma}{\bf}{\it}
\newenvironment\proposition{Proposition}{\bf}{\it}
\newenvironment\theorem{Theorem}{\bf}{\it}
\newenvironment\corollary{Corollary}{\bf}{\it}
\newenvironment\example{Example}{\bf}{\rm}
\newenvironment\problem{Problem}{\bf}{\rm}
\newenvironment\definition{Definition}{\bf}{\rm}
\newenvironment\remark{Remark}{\bf}{\rm}
\newenvironment\hypothesis{Hypothesis}{\bf}{\it}

%
%
\long\def\proof#1{\removelastskip\penalty-100\vskip\baselineskip\noindent{\bf
            Proof\if!#1!\else\ \ignorespaces#1\fi:\ }\ \ \ignorespaces}
\long\def\prf{\removelastskip\penalty-100\vskip\baselineskip\noindent{\bf
            Proof:\ }\ \ \ignorespaces}
\def\sq{\hbox{\rlap{$\sqcap$}$\sqcup$}}
\def\qed{\ifmmode\sq\else{\unskip\nobreak\hfil
           \penalty50\hskip1em\null\nobreak\hfil\sq
           \parfillskip=0pt\finalhyphendemerits=0\endgraf}\fi}
\def\endproof{\hfill\vrule height .6em width .6em depth
0pt\goodbreak\vskip.25in }
\def\eqn#1{\eqno{({\rm #1})}}
\def\+{\!+\!}
\def\-{\!-\!}
\def\={\ =\ }
\def\set#1#2{\big\{ \ #1\ \big|\ #2\ \big\}}
\def\eval#1{\big|\lower4pt\hbox{$\displaystyle\sst #1$}}

\null\vskip2.5truecm
%
%
\nopagenumbers
\centerline{\tafontt A Rigorous Analysis of the }
\centerline{\tafontt Superconducting Phase  }
\centerline{{\tafontt of an Electron - Phonon System}%
{\baselineskip=0.pt\parindent=.15in\footnote{$^{\clubsuit}$}
{\rm Lectures given by J. Magnen at the LXII les Houches
Summer School ``Fluctuating Geometries in Statistical Mechanics and in Field
Theory", August 2 -- September 9 1994.}}
}
\vskip20pt
\centerline{Joel Feldman{\baselineskip=0.pt\parindent=.15in\footnote{$^{*}$}
           { Research supported in part by
           the Natural Sciences and Engineering Research Council of Canada
            and the
           Schweizerischer Nationalfonds zur Forderung der wissenschaftlichen
           Forschung}}$^{\dagger}$}
\centerline{Department of Mathematics}
\centerline{University of British Columbia}
\centerline{Vancouver, B.C. }
\centerline{CANADA\ \   V6T 1Z2}\vskip.5truecm
\centerline{Jacques Magnen$^{\dagger}$,
Vincent Rivasseau{\baselineskip=0.pt\parindent=.15in\footnote{$^{\dagger}$}
          { Research supported in part by the Forschungsinstitut
          f\"ur Mathematik, Z\"urich}}}
\centerline{Centre de Physique Th\'eorique, CNRS UPR14}
\centerline{Ecole Polytechnique}
\centerline{F-91128 Palaiseau Cedex}
\centerline{FRANCE}\vskip.5truecm
\centerline{Eugene Trubowitz}
\centerline{Mathematik}
\centerline{ETH-Zentrum}
\centerline{CH-8092 Z\"urich}
\centerline{SWITZERLAND}\vskip2truecm

\vfill
\eject

\titleb{Introduction}

In these notes we give a very rough sketch of
nonperturbative methods of
many body quantum field theory that are powerful enough to rigorously control
weak coupling instabilities in condensed matter phyics, for example, the
Cooper instability in an electron - phonon system.

We restrict our attention to a
$ d\ge 2$ dimensional model for $\, \ell=0\, $ superconductivity.
Precisely, to the model
formally characterized by the Euclidean action
$$
{\cal A}(\psi,\bar\psi)\ =\ \int\!\!\dbar k
\, {\scriptstyle\big(ik_0\-e(\k)\big)}\bar\psi_{k}\psi_{k}
\ +\ {\textstyle{\la\over 2}}
\, \int\, {\textstyle\prod\limits_{{\scriptscriptstyle i=1}}
^{{\scriptscriptstyle 4}}}\, \dbar k_i
\ {\scriptstyle (2\pi)^{d\!+\!1}}
\delta{\scriptstyle \left(k_1\!+\!k_2\!-\!k_3\!-\!k_4\right)}\
\bar\psi_{k_1}\psi_{k_3}  {\scriptstyle \<k_1,k_2|{\rm V}|k_3,k_4\>}
\bar\psi_{k_2}\psi_{k_4}
$$
where  $\, e(\k) = {\k^2\over 2\m}-\mu\, $,
$\, \lambda > 0\, $. In these expressions, the electron fields are vectors
$\psi_k={\scriptscriptstyle\left(\matrix{\psi_{k,\uparrow}\cr
\psi_{k,\downarrow}\cr}\right)}$
and
$\bar\psi_k
={\scriptstyle\left(\bar\psi_{k,\uparrow},\bar\psi_{k,\downarrow}
\right)}$
whose components $\, \psi_{k,\sigma}\, ,
\,  \bar\psi_{k,\sigma}\, $ are
generators of an
infinite dimensional Grassmann algebra over $\bbbc.$ That is, the fields
anticommute with each other. The generating functional for the associated
connected, amputated Euclidean Green's functions is
$$
\cS(\phi,\bar\phi)\ =\ \log\,{\textstyle {1\over {\cal Z}}}
\, \int\,
e^{-{\cal A}(\psi+\phi,\bar\psi+\bar\phi)}
\, \smprod_{k,\si} d\psi_{k,\si}\,d\bar\psi_{k,\si}
$$

It is assumed that the  reduced interaction
$\, \<s^\prime,-s^\prime|{\rm V}|t^\prime,-t^\prime\>\, $ is
attractive and dominant
in the zero angular momentum sector. Here,
$\, k^\prime=(0,{\k\over |\k|}k_F)\, $ is the projection
of $\, k=(k_0,\k)\in \bbbr^{d+1}\, $ onto the Fermi surface.
In more detail, expanding in spherical harmonics,
$$
-\lambda\<s^\prime,-s^\prime|{\rm V}|t^\prime,-t^\prime\>
\ =\ \smsum\limits_{\ell \ge 0} \lambda_\ell{\scriptstyle(0)}
\, \pi_\ell(s^\prime,t^\prime)
$$
our assumption  becomes $\, \lambda_0{\scriptstyle(0)}>0\, $
and $\, \lambda_0{\scriptstyle(0)}\gg\lambda_\ell{\scriptstyle(0)}
\, ,\, \ell\ge 1\, $.

We now precisely formulate one of our goals. For each
$\, L>0\, $, let
$\, d\mu_L\, $ be the Grassmann Gaussian measure over the torus
$\, \bbbr^{d+1}/L\bbbz^{d+1}\, $ whose covariance is the
multiplication operator
$$
{1\over ik_0-e(\k)}\, (1-\delta_{k_0,0}\delta_{e(\k),0})
$$
on $\, \ell^2({2\pi\over L}\bbbz^{d+1})\, $. For convenience, let
$$
\int\dbar k f(k)\ =\ ({\textstyle{2\pi\over L}})^{d+1}\smsum_{k\in
{2\pi\over L}\bbbz^{d+1}} f(k)
$$
The finite volume action
$\, {\cal A}_{L,r}(\psi,\bar\psi)\, $ is given by
$$
{\cal A}_{L,r}(\psi,\bar\psi)
\ =\ {\cal A}(\psi,\bar\psi)
\ +\ \delta\mu(\lambda,\mu;L,r)\int\dbar k \bar\psi_k\psi_k
\ -\ r\int\dbar k (\bar\psi_{k\uparrow}\bar\psi_{-k\downarrow}
+\psi_{-k\downarrow}\psi_{k\uparrow})
$$
where there is a counterterm for the chemical potential and a small external
field that ultimately selects a pure phase. The corresponding
finite volume generating
functional is
$$
\cS_{L,r}(\phi,\bar\phi)\ =\ \log\,{\textstyle {1\over {\cal Z}}}
\, \int\,
e^{-{\cal A}_{L,r}(\psi+\phi,\bar\psi+\bar\phi)}
\, \smprod_{k,\si} d\psi_{k,\si}\,d\bar\psi_{k,\si}
$$
We are strongly convinced that the tools are at hand to give a completely
rigorous proof of the
\theorem{}{Let $\, d=2,3\, $ and let $\, \<k_1,k_2|{\rm V}|k_3,k_4\>\, $ be
a sufficiently regular, real function on $\, \bbbr^{4(d+1)}\, $ satisfying
$$\eqalign{
\<Rk_1,Rk_2|{\rm V}|Rk_3,Rk_4\>\ &=\ \<k_1,k_2|{\rm V}|k_3,k_4\>\cr
\<Tk_1,Tk_2|{\rm V}|Tk_3,Tk_4\>\ &=\ \<k_1,k_2|{\rm V}|k_3,k_4\>\cr
\<k_1,k_2|{\rm V}|k_3,k_4\>\ &=\ \<-k_3,k_2|{\rm V}|-k_1,k_4\>\cr
\ &=\ \<k_1,-k_4|{\rm V}|k_3,-k_2\>\cr
}
$$
for all $\, R\in O(d,\bbbr)\, $ where $\, Rk=(k_0,R\k)\, $, and
$\, Tk=(-k_0,\k)\, $. Let
$$
-\lambda\<s^\prime,-s^\prime|{\rm V}|t^\prime,-t^\prime\>
\ =\ \smsum\limits_{\ell \ge 0} \lambda_\ell{\scriptstyle(0)}
\, \pi_\ell(s^\prime,t^\prime)
$$
be the expansion of the rotation invariant reduced interaction in
spherical harmonics. Fix $\, \varepsilon>0\, $. Let $\, \lambda>0\, $ and
$\, \varepsilon^\prime>0\, $ be sufficiently small. If
$\, \lambda_0{\scriptstyle(0)}>0\, $
and $\, \varepsilon^\prime\lambda_0{\scriptstyle(0)}>
\lambda_\ell{\scriptstyle(0)}
\, ,\, \ell\ge 1\, $, then the limit
$$
\cS(\phi,\bar\phi)\ =\ \lim_{r\downarrow0}
\, \lim_{L\uparrow \infty}\, \cS_{L,r}(\phi,\bar\phi)
$$
exists and has the following properties:

\noindent
{\rm (i)} {\rm (}$U(1)$ Symmetry Breaking{\rm )}
There is a $\, \Delta>0\ $ with $\, \Delta={\sst\const}
e^{-{\const\over \lambda}}\, $ such that
$$
\<\psi_{k^\prime\uparrow}\psi_{-p\downarrow}\>
\ =\ \<\bar\psi_{-k^\prime\downarrow}\bar\psi_{p\uparrow}\>
\ =\ -{\textstyle {(2\pi)^{d+1}\over \Delta}}\, \delta(k^\prime-p)
$$

\noindent
{\rm (ii)} The $\, 2n\, $ point functions of $\, \cS(\phi,\bar\phi)\, $
with $\, n\, $ odd decay exponentially at a rate at least
$\, (1-\varepsilon)\Delta\, $.

\noindent
{\rm (iii)} {\rm (}Goldstone Boson{\rm )}
The $\, 2n\, $ point functions of $\, \cS(\phi,\bar\phi)\, $
with $\, n\, $ odd decay at least polynomially. In particular, there are
constants $\, c_1,c_2>0\, $ such that
$$
\int\dbar s\dbar t\dbar p\
\<\bar\psi_{s-q\uparrow}\bar\psi_{-s\downarrow}
+\psi_{-s+q\downarrow}\psi_{s\uparrow}\, ;\,
\bar\psi_{t-p\uparrow}\bar\psi_{-t\downarrow}
+\psi_{-t+p\downarrow}\psi_{t\uparrow}\>
\ =\
-{\textstyle{1\over c_1q_0^2+c_2\q^2}}\ +\ O(1)
$$
In other words, there is a channel in the four point function, due to the
$U(1)$ Goldstone boson, that does not decay exponentially fast.
}

For simplicity, we have stated a fragment of a more complete theorem.
\vskip.25truein

In order to systematically investigate the long range behavior
of correlation functions at low temperature, it is natural to use a
renormalization group analysis ([FT2],[FT3]) near the Fermi surface. This
entails slicing the free propagator around it singularity on the Fermi sphere.
The renormalization group generates an effective slice-dependent interaction.

To analyze the ultraviolet and infrared
behavior of a relativistic Euclidean
field theory, one defines a momentum $k$ to be
of scale $j$ if $|k|\approx 2^j$. Here, $2$ is just a fixed constant that
determines the scale units. As $j\rightarrow\infty$, the momentum $k$ approachs
the ultraviolet end of the model.  As $j\rightarrow-\infty$, $k$ approachs the
infrared end of the model.

In non-relativistic solid state physics the natural
scales consist of finer and finer shells around the Fermi surface.
For each negative integer
$j= 0, -1, -2,...$ the $j$-th slice contains all momenta in a shell of
thickness $2^{j}$ a distance $2^j$ from the singular locus
$$
\set{k\in\bbbr^{d+1}}{k_0=0,\ |\k|=\sqrt{2\m\mu}}
$$
The propagator for the
$j$-th slice is
$$
C^{j}(\xi_{1},\xi_{2})=\delta_{\sigma_{1},\sigma_{2}}\int
\dbar k{e^{i\<k,\xi_{1}-\xi_{2}\>_-}
\over ik_0-e(\k)}1_j\big(k_0^2+e(\k)^2\big)
\eqn{1}$$
where $\,1_j\big(k_0^2+e(\k)^2\big)\,$ is the characteristic function for
the set $\,2^j\le|ik_0-e(\k)|<2^{j+1}.$
For simplicity, we have introduced a sharp partition of unity even though a
smooth one is required for a complete, technically correct analysis [FT3,II.1].
Summing over $\, j\le 0\, $, we obtain the full infrared propagator
$\,
C(\xi_1,\xi_2) = \sum\limits_{j\le 0}
\, C_j(\xi_1,\xi_2)
\, $.
The full Schwinger functions are obtained by assigning each line of each
Feynman diagram a scale $j$ and then summing over all such assignments.

Each single scale propagator (1) is supported in momentum space
on a $d+1$ dimensional manifold with boundaries. The natural coordinates
for this
manifold are $k_0,\ \et~=~e(\k)$ and
$\k'=\sqrt{2\m\mu}\sfrac{\k}{|\k|}$. In these
coordinates the shell is $\set{k}{2^j\le\sqrt{k_0^2+\et^2}\le\const 2^j}$
and is topologically $S^{d-1} \times S^{1} \times [0,1]$.
However the first factor, the Fermi sphere $S^{d-1}$, should be viewed as
having a macroscopic radius of order 1 while the remaining factors
$S^{1}\times [0,1]$ should be viewed as having a small diameter of order
$2^{j}$ at scale $j$.

The fact that this manifold has two length scales, 1 and $2^j$, of
radically different size reflects the basic anisotropy between frequency
$k_0$ and momentum $\k$.
It implies, in contrast to the field theory case, that the behavior of
 $C^j(\xi_1,\xi_2)$ at large $\xi_1-\xi_2$ cannot be simply characterized
as `decay at rate $2^{-j}$'. Rather,  $C^j$ looks like
$$
\left|C^j(\xi_1,\xi_2)\right|\le\const 2^j \big[1+|\x_1-\x_2|\big]^{(1-d)/2}
\big[1+2^j|\xi_1-\xi_2|\big]^{-N}
$$
when a smooth cutoff function is used.

These shells induce an infrared renormalization group flow
that acts in the ladder approximation on the running coupling constants
$\, \lambda_\ell{\scriptstyle(j)}\, ,\,  \ell\ge 0\, $,
 associated to the quartic
local part  at scale $\, j\le 0\, $, by [FT3,I.85]
$$
\lambda_\ell{\scriptstyle(j-1)}\ =\ \lambda_\ell{\scriptstyle(j)}
+{\scriptstyle\beta(j)}\, \lambda_\ell{\scriptstyle(j)}^2
$$
where $\, {\scriptstyle\beta(j)}>0\, $ and
$\, \lim\limits_{j\rightarrow -\infty}{\scriptstyle\beta(j)}=\beta>0\, $.
In this approximation, our assumption  $\, \lambda_0{\scriptstyle(0)}>0\, $
and $\, \lambda_0{\scriptstyle(0)}>\lambda_\ell{\scriptstyle(0)}
\, ,\, \ell\ge 1\, $,
implies
$\, \lim\limits_{j\rightarrow -\infty}\lambda_0{\scriptstyle(j)}=\infty\, $
and that (see, [FT3]) to all orders in the full flow
$\, \lambda_0{\scriptstyle(j)}\, $ dominates
$\, \lambda_n{\scriptstyle(j)}\, ,n\ge 1\, $. In more detail,
$\, \lambda_0{\scriptstyle(j)}\, $ grows slowly
as $j$ goes down to $\, \delta+\const\, $
where,
$\, \delta=-[{1\over \lambda}]\, $ is the
symmetry breaking scale, and then quickly takes off to infinity.
The divergence of a flow
generated by a ``Fermi surface'' to a nontrivial fixed point is typical of
many
symmetry breaking
or mass generation phenomena in condensed matter physics.

This renormalization group analysis reveals three distinct energy regimes.
Fix $\, a\gg 2\, $ and let $\, \Delta\approx 2^\delta\, $ be
the BCS gap . In the first regime
at scales $\, j\, $, for which $\, 2^j>a\Delta\, $, the effective coupling
constant $\, \lambda_0(j)\, $ can be used as a small parameter. Symmetry
breaking takes place in the second regime where $\, {1\over a}\, \Delta
<2^j<a\Delta\, $. In the third regime, $\, 2^j<{1\over a}\, \Delta\, $,
the physics of the Goldstone boson dominates. As explained above, the effective
coupling constant is not small in the latter two regimes.
\vskip.25truein

\titleb{The ``First Regime'':}

In this section, we concentrate on the problem of summing high orders of
perturbation theory and , in particular, discuss the results of
[FMRT3]. To do so we first consider an artificial model
which retains
the essential difficulties we are interested in, but has no renormalization
problems. The toy world consists of
\item{-} $d+1$ dimensional Euclidean space-time
\item{-} four types of fermions, denoted $\psi_\uparrow,\ \psi_\downarrow,\
\bar\psi_\uparrow$ and $\bar\psi_\downarrow$, that play the roles of spin up
and spin down electrons and positrons/holes
\item{-} momenta ``morally'' in the range $M^j\le|p|\le M^{j+1}$ with $M>1$
being
some fixed constant. This is typical of one slice of a field theory. In a many
body
model  we would have $M^j\le|p_0|+\big||\p|-k_F\big|\le M^{j+1}$. In a
realistic model
we would have to sum over $j$ using the renormalization group.

\noindent We say ``morally'' because momentum is never actually going to
appear in the toy world. Instead we are going to mimic the assumed momentum
range by two space-time properties of the model. First, because the momentum
space of
our toy world has volume $M^{j(d+1)}$ the Pauli exclusion principle says
that there can be at most one $\psi_\uparrow$, for example, in any region
of volume $M^{-j(d+1)}$ in position space.
Thus we define the fields of our model to be
$$
\left\{\psi_\uparrow(x),\psi_\downarrow(x),\bar\psi_\uparrow(x),
\bar\psi_\downarrow(x)\ \big|\ x\in W:\hskip-.5pt=M^{-j}\bbbz^{d+1}\right\}
$$
They are the generators of a Grassmann algebra. Thus
$$
\optbar \psi_\al(x)\optbar \psi_\be(y)
=-\optbar \psi_\be(y)\optbar \psi_\al(x)
$$
and in particular
$$
\left(\optbar \psi_\al(x)\right)^2=0
$$

The second concerns the propagator. That is, the free two point
Euclidean Green's function. The interacting two point Euclidean Green's
function is
$$
S_2(x,x')={\int \psi_\uparrow(x)\bar\psi_\uparrow(x')e^{-\la V}d\mu_C\over
\int e^{-\la V}d\mu_C}
$$
where the interaction
$$
V={1\over 2}\sum_{y\in W}M^{-j(d+1)}\bar\psi_\uparrow(y)\bar\psi_\downarrow(y)
\psi_\downarrow(y)\psi_\uparrow(y)
$$
and
$$
d\mu_C=\exp\bigg\{\sum_{z,z'\in W\atop\si\in\{\uparrow,\downarrow\}}
\bar\psi_\si(z')C^{-1}(z',z)\psi_\si(z)\bigg\}
\prod_{z\in W\atop\si\in\{\uparrow,\downarrow\}}
d\psi_\si(z)d\bar\psi_\si(z)
$$
is the Grassmann Gaussian measure with covariance $C$, to be specified shortly.

Here are all the properties of
Grassmann Gaussian measures that we are going to use. The symbol
$\int\ \cdot\ d\mu_C$ is a linear functional that assigns a complex number
to every polynomial in the fields and that obeys
$$\eqalign{
1.\ \ \ \int \psi_\si(x)\bar\psi_{\si'}(y)d\mu_C&\=\de_{\si,\si'}C(x,y)\cr
2.\  \int \psi_\si(x)F(\psi,\bar\psi)d\mu_C&\=
\sum_{y\in M^{-j}\bbbz^{d+1}}C(x,y)\int{\de\hfill\over\de\bar\psi_\si(y)}
F(\psi,\bar\psi)d\mu_C\cr
\dst \int F(\psi,\bar\psi)\bar\psi_\si(y)d\mu_C&\=
\sum_{x\in M^{-j}\bbbz^{d+1}}\int
F(\psi,\bar\psi){\de\hfill\over\de\psi_\si(x)}d\mu_C\ C(x,y)\cr
&\=\sum_{x\in M^{-j}\bbbz^{d+1}}C(x,y)\int{\de\hfill\over\de\psi_\si(x)}
F(\psi,\bar\psi)d\mu_C \cr
}$$
Except for signs the left and right derviatives ${\de\hfill\over\de\psi_\si(y)}
F(\psi,\bar\psi)$ and $F(\psi,\bar\psi){\de\hfill\over\de\psi_\si(y)}$
behave like ordinary derivatives.

We assume, as the second characteristic of our momentum range, that the
covariance $C$
decays at a rate typical of a smooth function whose Fourier transform has
support in a
neighbourhood of $|p|=M^j$. Precisely,
$$
|C(x,y)|\le \const M^{(d+1)j/2}e^{-M^j|x-y|}\ .
$$
The coefficient $M^{(d+1)j/2}$ is chosen to give power counting typical of a
strictly renormalizable field theory. The position space behaviour of
the many-Fermion propagator is somewhat more complicated than this. However,
by decomposing the Fermi surface into a union of $M^{-(d-1)j}$ ``rectangles''
of side $M^j$,  one can think of the many-Fermion field at scale $j$ as a sum
$
\psi^{(j)}=\sum_{\al}\psi^{(j,\al)}
$
of $M^{-(d-1)j}$ independent fields with
each ``coloured'' field having a covariance that obeys
$
|C(x,y)|\le \const M^{dj}e^{-M^j|x-y|}
$. See [FMRT3].
\theorem{}{ Let $S_{2,n}(x,x')$ be the coefficient of $\la^n$ in the
formal power series expansion of $S_2(x,x')$. That is,
$S_2(x,x')=\sum_{n=0}^\infty S_{2,n}(x,x')\la^n$. There exists
a constant $R$, independent of $j,x,x'$, such that
$$
\sup_x\sum_{x'}|S_{2,n}(x,x')|\le K_jR^n\ .
$$
In other words $S_2$ is analytic in $|\la|<{1\over R}$. In other words, the
sum of all connected Feynman diagrams converges for all $|\la|<{1\over R}$.}
\prf
We first describe the logic of the proof. Denote by $S_2(x,x';\La)$
the two point function of the model gotten by restricting the world to a
finite subset of $\La$ of $W$. It is easy to see [FMRT3], by Gram's
or Hadamard's inequality, that both the numerator and denominator of
$S_2(x,x';\La)$ are entire functions of $\lambda$. The denominator
$\int e^{-\la V}d\mu_C$ can have many  $\Lambda$ dependent zeros. But when
$\lambda=0$, the denominator is one so that $S_2(x,x';\La)$ is meromorphic
on all of $\bbbc$ and analytic at zero. We shall develop a formal power
series expansion for $S_2(x,x';\La)$ with the property that for every $N$
$$
S_2(x,x';\La)=\sum_{n=0}^N S_{2,n}(x,x';\La)\lambda^n+
O\left(\lambda^{N+1}\right).
$$
A priori we do not claim that the tail
$O\left(\lambda^{N+1}\right)$ is uniform in $\Lambda$.
Nevertheless, since $S_2(x,x';\La)$ is analytic at zero we must have
$$
S_2(x,x';\La)=\sum_{n=0}^\infty S_{2,n}(x,x';\La)\lambda^n
\eqn{*}$$
in some, possibly $\Lambda$ dependent, neighborhood of
zero. We remark in passing that $S_{2,n}(x,x';\La)$ must be the sum of all
connected Feynman diagrams of order $n$ with $2$ external legs, since
we have  an asymptotic expansion.

The heart of the proof is to show that there exists a constant $R$,
independent of $\La,j$ and a constant $K_j$ independent of $\La$ such that
$$
\sup_x\sum_{x'}|S_{2,n}(x,x';\La)|\le K_jR^n
\eqn{**}$$
As a consequence, equation ($\ast$) applies for all
$|\lambda|<R^{-1}$.  Any zeroes of the denominator that appear in
this disk must be cancelled by zeroes of the numerator. It shall also
be clear from the proof of ($\ast$$\ast$) that the limits
$S_{2,n}(x,x')=\lim\limits_{\Lambda\rightarrow W}
S_{2,n}(x,x';\La)$ exist.  This will prove, by the Lebesgue dominated
convergence theorem, that
$$
S_{2}(x,x')=\lim_{\Lambda\rightarrow W}S_{2}(x,x';\La)
=\sum_{n=0}^\infty S_{2,n}(x,x')\lambda^n
$$
for all $|\lambda|<R^{-1}$ and that the coefficients $S_{2,n}(x,x')$
obey the bound ($\ast\ast$).

We now describe the expansion used. To emphasize that everything is uniform
in $\La$, we supress $\La$. The first step is to use integration by
parts (Property 2) to turn the $\psi_\uparrow(x)$ of the two point
function into a covariance:
$$\eqalign{
S_2(x,x')&={\int \psi_\uparrow(x)\bar\psi_\uparrow(x')e^{-\la V}d\mu_C\over
\int e^{-\la V}d\mu_C}\cr
&=C(x,x')+{\sum_y\la C(x,y){\dst \int} \bar\psi_\uparrow(x')
{\de\ \ V\hfill\over\de\sbar\psi_\uparrow(y)}
e^{-\la V}d\mu_C\over\int e^{-\la V}d\mu_C}\cr
}$$
The first term is the trivial Feynman diagram giving the free value of $S_2$.
For  the second, apply integration by parts again to turn the
$\bar\psi_\uparrow(x')$ into another propagator.
$$\eqalign{
S_2(x,x')&=C(x,x')-{\sum_{y,y'}\la^2 C(x,y)C(y',x'){\dst \int}
\Big[{\de\hfill\over\de\psi_\uparrow(y')}
                             {\de\hfill\over\de\sbar\psi_\uparrow(y)}V\Big]
e^{-\la V}d\mu_C\over\int e^{-\la V}d\mu_C}\cr
&\hskip.8in-{\sum_{y,y'}\la^2 C(x,y)C(y',x'){\dst \int}
{\de\ \ V\hfill\over\de\sbar\psi_\uparrow(y)}
{\de\ \ V\hfill\over\de\psi_\uparrow(y')}
e^{-\la V}d\mu_C\over\int e^{-\la V}d\mu_C}\cr
}$$
In each step select any $\optbar\psi$ downstairs and use integration by parts
to
turn it into one end of a propagator. When a term has no fields downstairs, the
$\int e^{-\la V}d\mu_C$ in the numerator exactly cancels that in the
denominator, leaving a Feynman diagram. This was how the trivial diagram $C$
arose. Leave such terms alone. Upon completion of the expansion, we have
$S_2(x,x')$ expressed as the sum of all connected two point Feynman diagrams.

To illustrate the principal difficulty in estimating $S_2$ consider the
following $n^{\rm th}$ order term that arises in the midst of the
expansion:\hfill\break
$$\eqalign{
&{\la^n\over 2^n}\sum_{y_1,\cdots,y_n\in W}M^{-j(d+1)n}
C(x,y_1)C(y_1,y_2)\cdots C(y_n,x'){\int\prod_{m=1}^n\bar\psi_\downarrow(y_m)
\psi_\downarrow(y_m)e^{-\la V}d\mu_C\over\int e^{-\la V}d\mu_C}\cr
}$$
The functional integral
$$\eqalign{
&\int\prod_{m=1}^n\bar\psi_\downarrow(y_m)
\psi_\downarrow(y_m)e^{-\la V}d\mu_C\cr
&\hskip.5pt=-\sum\limits_{z\in W}C(z,y_1)
\int{\de\hfill\over\de\psi_\downarrow(z)}\left[\psi_\downarrow(y_1)
\prod\limits_{m=2}^n\bar\psi_\downarrow(y_m)\psi_\downarrow(y_m)
                            e^{-\la V}\right]d\mu_C\cr
&\hskip.5pt=-\sum_{i=1}^nC(y_i,y_1)\int \psi_\downarrow(y_1)\cdots
\psi_\downarrow(y_i)\hskip-28pt\vbox{\hrule height 3pt width 28pt depth-2.5pt}
\cdots\bar\psi_\downarrow(y_n)
\psi_\downarrow(y_n)e^{-\la V} d\mu_C+O(\la)\cr
}$$
We did a single integration by parts to get rid of $\bar\psi_\downarrow(y_1)$
and ended up with $n$ terms of order $\la^n$. If we perform $n-1$ further
integrations by parts to get rid of $\bar\psi_\downarrow(y_2),\cdots,
\bar\psi_\downarrow(y_2)$ we will generate $n!$ diagrams of order $\la^n$.
Naive bounds on these $n!$ terms will fail to produce an acceptable bound on
$S_{2,n}.$

Fortunately, the Pauli exclusion principle saves us. Note first that,
if $y_m~=~y_{m'}$ for any $m\ne m'$, then
$\psi_\downarrow(y_m)\psi_\downarrow(y_m')
=-\psi_\downarrow(y_m')\psi_\downarrow(y_m)$ so that
$\psi_\downarrow(y_m)\psi_\downarrow(y_m')=0$ and hence
$\psi_\downarrow(y_1)\prod\limits_{m=2}^n\bar\psi_\downarrow(y_m)
                                                 \psi_\downarrow(y_m)=0.$
Let $A_i=\int \psi_\downarrow(y_1)\cdots
\psi_\downarrow(y_i)\hskip-28pt\vbox{\hrule height 3pt width 28pt depth-2.5pt}
\cdots\bar\psi_\downarrow(y_n)
\psi_\downarrow(y_n)e^{-\la V} d\mu_C$. Then we may bound
$$\eqalignno{
\sum_{i=1}^n\left|C(y_i,y_1)A_i\right|
&\le\max_{1\le i\le n}\left|e^{M^j|y_1-y_i|/2}C(y_i,y_1)A_i \right|
                                \sum_{i=1}^ne^{-M^j|y_1-y_i|/2}\cr
&\le\max_{1\le i\le n}\left|M^{j(d+1)/2}e^{-M^j|y_1-y_i|/2}A_i \right|
                  \sum_{y\in M^{-j}\bbbz^{d+1}}e^{-M^j|y_1-y|/2}\cr
&=\max_{1\le i\le n}\left|M^{j(d+1)/2}e^{-M^j|y_1-y_i|/2}A_i \right|
                  \sum_{x\in \bbbz^{d+1}}e^{-|x|/2}\cr
&=\cE\max_{1\le i\le n}\left|M^{j(d+1)/2}e^{-M^j|y_1-y_i|/2}A_i \right|
&{\rm (***)}}$$
where $\cE=\sum_{x\in \bbbz^{d+1}}^ne^{-|x|/2}<\infty$.
The crucial consequence of the Pauli exclusion principle, that the $y_i$'s all
are different, was used in going from line one to line two. Think of
$M^{j(d+1)/2}e^{-M^j|y_1-y_i|/2}$ as a propagator (replacing $C(y_1,y_i)$)
for a line in a graph.
This propagator joins a vertex at $y_1$ to a vertex at $y_i$. The fields
$\optbar \psi$ downstairs in the functional integral $A_i$ are
external legs for the graph.

Proceed by induction. In each step of the induction we integrate by
parts once and apply the above bounding procedure. We start the
$k^{\rm th}$ step in the induction process with a maximum over $k-1$
indices like the $i$ in ($\ast\ast\ast$) and we end the step with $k$ such
indices.
We think of each such index as specifying the target vertex of a propagator.
At the end of the expansion we find
$$\eqalign{
|S_{2,n}(x,x')|&\le{1\over 2^n}\max_G\sum_{y_1,\cdots,y_n}M^{-j(d+1)n}
\prod_{\ell\in
G}\left[(\cE+1)M^{j(d+1)/2}e^{-M^j|y_{i_\ell}-y_{f_\ell}|/2}\right]\cr
}$$
The maximum is over all connected Feynman digrams with two one-legged vertices,
labeled $x,x'$ and $n$ four-legged vertices labeled $y_1,\cdots,y_n$. The
labels
of the two vertices at the ends of line $\ell$ are denoted $i_\ell$ and
$f_\ell$.
The reason for the $+1$ in $\cE+1$ is that each functional derivative arising
from an
application of the integration by parts formulae can act on the exponent as
well as
on interaction vertices downstairs.
In preparation  for bounding the graph $G$, select a spanning tree $T$ for $G$.
A spanning tree is a subgraph $T\subset G$ which has no loops and contains all
the
vertices of $G$. Bound all factors $e^{-M^j|y_{i_\ell}-y_{f_\ell}|/2}$ that are
associated with lines $\ell\in G\setminus T$ by one. Then apply
$$
\sum_{y\in W} e^{-M^j|y'-y|/2}\le \cE
$$
to each vertex of $G$ starting with those farthest from $x$ in the partial
ordering of $T$. The result is
$$\eqalign{
\sum_{x'}|S_{2,n}(x,x')|&\le{1\over
2^n}\max_G\sum_{y_1,\cdots,y_n,x'}M^{-j(d+1)n}
\prod_{\ell\in
G}\left[(\cE+1)M^{j(d+1)/2}e^{-M^j|y_{i_\ell}-y_{f_\ell}|/2}\right]\cr
&\le{1\over
2^n}M^{-j(d+1)n}\max_G(\cE+1)^{|G|}M^{|G|j(d+1)/2}\sum_{y_1,\cdots,y_n,x'}
\prod_{\ell\in T}\left[e^{-M^j|y_{i_\ell}-y_{f_\ell}|/2}\right]\cr
&\le{1\over 2^n}M^{-j(d+1)n}\max_G(\cE+1)^{|G|}M^{|G|j(d+1)/2}\cE^{n+1}\cr
}$$

As we are currently considering an $n^{\rm th}$ order diagram contributing to
the
two point function
$$
|G|={2+4n\over 2}=2n+1
$$
and the final bound is
$$\eqalign{
\sum_{x'}|S_{2,n}(x,x')|&\le{1\over 2^n}
M^{-j(d+1)n}(\cE+1)^{2n+1}M^{(2n+1)j(d+1)/2}
\cE^{n+1}\cr
&\le {(\cE+1)^{3n+2}\over 2^n} M^{j(d+1)/2}\ ,
}$$
which proves the Theorem with $R={1\over 2} (\cE+1)^3$ and
$K_j=(\cE+1)^2M^{j(d+1)/2}$.

More generally, for a $p$ point function, $|G|={p+4n\over 2}=2n+p/2$, the
number of
sums controlled by the tree decay is $n+p-1$ and
$$\eqalign{
\sum_{x_2,\cdots,x_p}|S_{p,n}(x_1,\cdots,x_p)|
&\le{1\over 2^n} M^{-j(d+1)n}(\cE+1)^{2n+p/2}M^{(2n+p/2)j(d+1)/2}
\cE^{n+p-1}\cr
&\le {(\cE+1)^{3n+3p/2-1}\over 2^n} M^{pj(d+1)/4}\ .
}$$
\endproof
\vskip.25truein

The preceding techniques for summing perturbation
theory do not apply directly to many fermions systems. The assumption
$M^j\le|p|\le M^{j+1}$ is an over simplification.
The presence of the Fermi surface forces
$M^j\le|p_0|+\big||\p|-k_F\big|\le M^{j+1}$ and this makes it much more
difficult to implement the Pauli exclusion principle quantitatively.

To explain the difficulty, observe that the shell in momentum space about
the Fermi surface has volume $\, M^{2j}\, $, while the position space
volume of a ``dual'' box is $\, M^{-(d+1)j}\, $. The Pauli exclusion principle
now permits a dual box to contain
, in contrast to the ``toy'' system discussed above,
$\, O( M^{-(d-1)j})\, $ electrons. For $\, d=1\, $ there is no true Fermi
surface and there are $\, O(1)\, $ electrons in a dual box. As the dimension
grows the Pauli principle becomes progressively weaker.

There are three naive ways to force the volume in phase space to be
independent of the scale $\, j\, $. One either makes the dual boxes smaller,
or decomposes the shell into sufficiently small sectors, or both. In each case,
the number of electrons in such a constrained region would be of order one,
achieving duality in the sense of the exclusion principle.
The first alternative, however, violates duality in the sense of decay of the
propagator.

Let us decompose
the shell $\, \big||\k|-k_F\big|\approx M^j\, $ about the Fermi surface
into
$M^{-(d-1)j}$ sectors of side $M^j$ by another partition of unity. The
free propagator at scale $j$ in sector ${\scriptstyle \Sigma}$ is given
by
$$
C_{j,\Sigma}(\xi_1,\xi_2)\ =\ \delta_{\sigma_1,\sigma_2}
\, \int_{\bbbr^{d\!+\!1}}\, \dbar k
\, {e^{i\<k,\xi_1\!-\!\xi_2\>_-}\over ik_0-e(\k)}
\, 1_j\big(k_0^2+e(\k)^2\big)\, S_\Sigma(\k)
$$
where $S_\Sigma$ is supported on the sector ${\scriptstyle \Sigma}$. Of course,
$\,C_j=\sum\limits_ {\scriptstyle \Sigma}C_{j,\Sigma}\, $. There is a
corresponding decomposition of the fields. Observe that there are at most
two spin one half fields at scale $j$ in a position space box of
side $M^{-j}$ and momenta in the support of
$\, 1_j\big(k_0^2+e(\k)^2\big)\, S_\Sigma(\k)\, $, while there are
$O(M^{-(d-1)j})$ fields with momenta in the whole shell. That is, sectors
enforce the full Pauli exclusion principle, while the whole shell allows a
scale dependent accumulation of fields, reminiscent of Bosons.

It is easy to derive standard momentum space
power counting for an individual
graph using sectors.
As usual, one selects a spanning tree for the graph. To each line not in the
tree there is a corresponding momentum loop obtained by joining its ends
through a path in the tree. This construction produces a  complete set of
independent loops. Ignoring unimportant constants, each propagator is
bounded by it supremum $\, M^{-j}\, $. The volume of integration for each
loop is now $\, M^{(d+1)j}\, $. A priori, there is one sector sum with
$\, M^{-(d-1)j}\, $ terms for each line. But, by conservation of momentum,
there is only one sector sum per loop and one obtains the usual
$\, M^{{1\over 2}(4-E)j}\, $ where $\, E\, $ is the number of external lines.

In the course of a non-perturbative construction, estimates cannot be made
graph by graph because there are too many of them.
Rather, as in the proof of the theorem, the perturbation series must be blocked
and the blocks estimated as units. The blocks are estimated using the
exclusion principle to implement strong cancellations between the
roughly $\, n!^2\, $ graphs of order $\, n\, $. However, once the series is
blocked, momentum loops can't be defined and the argument leading to the power
counting estimate cannot be made. Conservation of momentum has to be
implemented at each vertex rather than through loops. Loosely speaking, the
Fermi surface makes it hard to fit
the Pauli exclusion principle and conservation of momentum together.

Specializing to two dimensions, one can show (using the observation that
four planar vectors of equal length whose sum is zero form a parallelogram,
see, [FMRT3]) that the number
of active sector 4-tuples at a vertex is of order $\, |j|M^{-2j}\, $. The
factor $\, M^{-2j}\, $ is natural since a parallelogram is determined by two
of its sides. the logarithmic factor arises from the degenerate situation
in which all four vectors are roughly collinear. One can
combine [FMRT3] the methods used to prove the theorem with the
decomposition of the propagator into sectors to obtain a rigorous
nonperturbative analysis in two dimensions of the full many electron system
down to the scale $\, a\Delta\, $, that is throughout the first regime.

In three dimensions, the parallelogram is hinged and the logarithm $\, |j|\, $
jumps to the power $\, M^{-{1\over 2}j}\, $. This power makes the
problem of controlling the first regime in three dimensions much more
difficult. New ideas are required that we will not discuss here.

Anderson [A] has conjectured the failure of Fermi liquid theory in
two dimensional interacting electron systems even at weak coupling. He
argues that an {\it orthogonality catastrophe} leads to Luttinger liquid
behavior roughly
characterized by the continuity of the number density $\, n_k\, $ across the
Fermi surface.
Our two dimensional expansion can also be used to construct weakly coupled
electron systems that in contrast to Anderson's conjecture
are rigorously Fermi liquids
(see, [FKLT]) in the sense that $\, n_k\, $ jumps down discontinuously
at the Fermi surface.
\vskip.25truein

\titleb{The ``Second Regime'':}

We now make an Ansatz that can be rigorously justified.
We suppose that the effective vertex at scale
$\, \delta_{\scriptscriptstyle +}\!=\delta+\const\, $ is
the BCS interaction for Cooper pairs
$$
V_{\rm eff}\ =\ -2\lambda_0{\scriptstyle(\delta_{\scriptscriptstyle +}\!)}
\int_{|q|<\const \Delta}\dbar q\dbar t\dbar s
\ \bar\psi_{t+{q\over 2},\uparrow}
\bar\psi_{-t+{q\over 2},\downarrow}
\psi_{-s+{q\over 2},\downarrow}
\psi_{s+{q\over 2},\uparrow}
$$
Here, $\, 0<\lambda_0{\scriptstyle(\delta^\prime)}=O(1)\, $,
$\, \Delta\approx M^\delta\, $ the symmetry breaking energy (the BCS gap),
and all the fields are at scale
$ \delta_{\scriptscriptstyle +} $ so that the integrals are implicitly
constrained by
$\, {\scriptstyle |e(\pm t+{q\over 2})|}
\, ,\, {\scriptstyle |e(\pm s+{q\over 2})|}
\, ,\, {\scriptstyle |t_0|}\, ,\, {\scriptstyle |s_0|}<\const \Delta\, $.
Approximating $\pm k+{q\over 2}$ by $\pm k$, using the notation
$V=M^{-(d+1)\delta}$ and
introducing sectors at scale $\delta_{\scriptscriptstyle +} $
, the effective vertex becomes
$$
\sum_{\Sigma_1,\Sigma_2}
-{2\lambda_0{\scriptstyle(\delta_{\scriptscriptstyle +}\!)}
\over V}
\int\dbar s_1\dbar s_2
\ \bar\psi_{\Sigma_1,\uparrow}{\scriptstyle(s_1)}
\bar\psi_{-\Sigma_1,\downarrow}{\scriptstyle(s_1)}
\psi_{-\Sigma_2,\downarrow}{\scriptstyle(s_2)}
\psi_{\Sigma_2,\uparrow}{\scriptstyle(s_2)}
$$
We denote by $-{\scriptstyle\Sigma}$ the sector antipodal
to ${\scriptstyle\Sigma}$. Note that the sums run
individually over $(\const\Delta)^{-(d-1)}$ sectors.

Our vertex has the structure typical of an
$N(=(\const\Delta)^{-(d\-1)})-$component
vector model.  Pictorially, sectors (``colors'')
${\scriptstyle\Sigma_1}$ and $-{\scriptstyle\Sigma_1}$ enter at the left
of an interaction squiggle and the sectors ${\scriptstyle\Sigma_2}$
and $-{\scriptstyle\Sigma_2}$ at the right. Thus, the sector is
conserved  up to a flip as it flows along
particle lines in a graph. We now check, by rough power counting that the
effective coupling constant of the vertex is $1/N$.

To estimate the size of the vertex, observe that the momentum space
free propagator
at scale $\delta_{\scriptscriptstyle +} $ in
sector ${\scriptstyle\Sigma}$ is
$$
{\textstyle
{ 1_{\, \delta_{\scriptscriptstyle +}}\!\big(k_0^2+e(\k)^2\big)\, S_\Sigma(\k)
\over ik_0-e(\k)}
}
\ \approx\
{\textstyle
{ 1_{\, \delta_{\scriptscriptstyle +}}\!\big(k_0^2+e(\k)^2\big)
\, S_\Sigma(\k)\over \Delta}
}
$$
Fixing $\, {\scriptstyle \Sigma_1}\, $ and $\, {\scriptstyle \Sigma_2}\, $
there is
one $\, d+1\, $ dimensional momentum integration and two
propagators per vertex.
Thus,
the power counting weight of the vertex is
$$
\int \dbar k\, {\textstyle
{ 1_{\, \delta_{\scriptscriptstyle +}}\!\big(k_0^2+e(\k)^2\big)\, S_\Sigma(\k)
\over \Delta^2}
}
\ \approx\ {\textstyle {\Delta^{d+1}\over \Delta^2}}\ =\ \Delta^{d-1}
\ =\ O(1/N)
$$

Recall that diagramatically, the
Goldstone boson
propagator is the sum of all one particle irreducible four legged
graphs. The simplest subseries is the geometric series of bubbles. It follows
from the last two paragraphs that the
bubble is $O(1)$ in $1/N$ and all other four legged graphs are of higher
order.
It follows from these remarks that the full Goldstone boson
propagator can be expanded in powers of $1/N$ around the explicit sum
of all bubbles. The large number of components can be used to
rigorously control the nonperturbative effects of
pairing and the associated $U(1)$ Goldstone boson that appear in the transition
between the first and second regimes.

\titleb{The ``Third Regime'':}

In the third regime the electrons have already paired and the number symmetry
is broken. The physics is
dominated by the Golstone boson that mediates the interaction between
pairs of Cooper pairs. The Goldstone boson will appear as a
singularity in the four point function. That is,
$$
\int\dbar s\dbar t\dbar p\
\<\bar\psi_{s-q\uparrow}\bar\psi_{-s\downarrow}
+\psi_{-s+q\downarrow}\psi_{s\uparrow}\, ;\,
\bar\psi_{t-p\uparrow}\bar\psi_{-t\downarrow}
+\psi_{-t+p\downarrow}\psi_{t\uparrow}\>
\ =\
-{\textstyle{1\over c_1q_0^2+c_2\q^2}}\ +\ O(1)
$$

To rigorously control
the  symmetry breaking we study the
 effective interaction
\def\cVf{{\cal V}_{\rm eff}}
$$\eqalign{
\cVf &=\ -2g^2\, \int\, \dbar s\, \dbar t\, \dbar q\
\, \bar\psi_\uparrow{\scriptstyle(t+{q\over 2})}
\, \bar\psi_\downarrow{\scriptstyle(-t+{q\over 2})}
\, \psi_\downarrow{\scriptstyle(-s+{q\over 2})}
\, \psi_\uparrow{\scriptstyle(s+{q\over 2})}\cr
\ &=\ -2g^2\, \int\, \dbar p \dbar q
\, \left(\int\dbar t\, \bar\psi_\uparrow{\scriptstyle(t+{p\over 2})}
\, \bar\psi_\downarrow{\scriptstyle(-t+{p\over 2})}\right)
\, B(p,-q)
\, \left(\int\dbar s\, \psi_\downarrow{\scriptstyle(-s+{q\over 2})}
\, \psi_\uparrow{\scriptstyle(s+{q\over 2})}\right)\cr
}
$$
with
$\
B(p,q)\ =\ {\scriptstyle (2\pi)^{\scriptscriptstyle d+1}}\delta(p\!+\!q).
$
Note that, by antisymmetry,
$$
 \int\, \dbar s\, \dbar t\, \dbar q\
\, \bar\psi_\uparrow{\scriptstyle(t+{q\over 2})}
\, \bar\psi_\uparrow{\scriptstyle(-t+{q\over 2})}
\, \psi_\uparrow{\scriptstyle(-s+{q\over 2})}
\, \psi_\uparrow{\scriptstyle(s+{q\over 2})}=0
$$
\vskip.15truein

Let $\ (\gamma_1,\gamma_2)\ $ be a $\, \bbbc^2\, $ valued
Gaussian variable with the real, even
covariance
$$
\<\gamma_i(p)\gamma_j(q)\>\ =\ \delta_{i,j}\,  B(p,q)
$$
Observe that the position space covariance
$\ \<\gamma_i(\xi_1)\gamma_j(\xi_2)\>=\de_{i,j}\de(\xi_1-\xi_2)\ $ is also
real.
Thus we can choose $\ga_i(\xi)$ to be real valued. Set
$$
\Ga(\xi)\ =\ \gamma_1(\xi)-i\gamma_2(\xi)
$$
We have
$$
e^{-\cVf}\ =\ \int
e^{g\int\!\! \dbar t\dbar q\left(\Ga(q)
\bar\psi_\uparrow{(t+{q\over 2})}\bar\psi_\downarrow{(-t+{q\over 2})}
+\overline{\Ga}(q)\psi_\downarrow{(-s+{q\over 2})}\psi_\uparrow{(s+{q\over 2})}
\right)}d\mu{(\gamma_1,\gamma_2)}
$$
since for all functions $\, X(q)\, $ and $\, Y(q)\, $
$$\eqalign{
\int & e^{\int\!\! \dbar q\, (X(q)\Ga(q)+Y(q)\overline{\Ga}(q))}
d\mu{(\gamma_1,\gamma_2)}
= \int e^{\int\!\! \dbar q \big(X(q)+Y(-q)\big)
\gamma_1(q)-i\big(X(q)-Y(-q)\big)\gamma_2(q)}
d\mu{(\gamma_1,\gamma_2)}\cr
&\quad=\ e^{{1\over 2}\int\!\!\dbar p\dbar q \big(X(p)+Y(-p)\big)
B(p,q)\big(X(q)+Y(-q)\big)}
e^{-{1\over 2}\int\!\!\dbar p\dbar q \big(X(p)-Y(-p)\big)
B(p,q)\big(X(q)-Y(-q)\big)}\cr
&\quad=\ e^{2\int\!\!\dbar p\dbar q\,  X(p)B(p,q)Y(-q)}\cr
}
$$

Changing variables and combining terms,
$$\eqalign{
&g\int\!\! \dbar t\dbar q\left(\Ga(q)
\bar\psi_\uparrow{(t+\sfrac{q}{2})}\bar\psi_\downarrow{(-t+\sfrac{q}{2})}
+\overline{\Ga}(q)\psi_\downarrow{(-s+\sfrac{q}{2})}
\psi_\uparrow{(s+\sfrac{q}{2})}\right)\cr
&\=g\int\!\! \dbar t\dbar q\left(\Ga(q)
\bar\psi_\uparrow{(t+\sfrac{q}{2})}\bar\psi_\downarrow{(-t+\sfrac{q}{2})}
+\overline{\Ga}(-q)\psi_\downarrow{(-t-\sfrac{q}{2})}
\psi_\uparrow{(t-\sfrac{q}{2})}\right)\cr
&\= g\int\dbar q\, \dbar t\,
\left(\matrix{\bar\psi_\uparrow{(t+\sfrac{q}{2})}
&\psi_\downarrow{(-t-\sfrac{q}{2})}\cr}\right)
\left(\matrix{0&\Ga(q)\cr
\overline{\Ga}(-q)&0\cr}\right)
\left(\matrix{\psi_\uparrow{(t-\sfrac{q}{2})}
\cr\bar\psi_\downarrow{(-t+\sfrac{q}{2})}\cr}\right)
\cr
&\= g\int\dbar q\, \dbar t\ \bar\bpsi{(t+\sfrac{q}{2})}
\, {\tst\left(\matrix{0&{\Ga(q)}\cr
{ \overline{\Ga}(-q)}&0\cr}\right)}
\, \bpsi{(t-\sfrac{q}{2})}
\= g\int d\xi\ \bar\bpsi{(\xi)}
\, {\tst\left(\matrix{0&{\Ga(\xi)}\cr
{\overline{\Ga}(\xi)}&0\cr}\right)}
\, \bpsi{(\xi)}
\cr
}
$$
where
$$\eqalign{
\bpsi(k)&=\left(\matrix{\bpsi^1(k)\cr
                           \bpsi^2(k)}\right)
=\left(\matrix{\psi_{k\uparrow}\cr
                           \bar\psi_{-k\downarrow}}\right)\cr
\bar\bpsi(k)&=\left(\matrix{\bar\bpsi_1(k)&
                           \bar\bpsi_2(k)}\right)
=\left(\matrix{\bar\psi_{k\uparrow}&
                           \psi_{-k\downarrow}}\right)\cr
}$$
are Nambu fields.
For convenience set
$$
\gamma\ =\ \gamma_1 \sigma^1+\gamma_2 \sigma^2
\ =\ {\tst\left(\matrix{0&{\Ga(\xi)}\cr
{\overline{\Ga}(\xi)}&0\cr}\right)}
$$
Then,
$$
e^{-\cVf}\ =\ \int
\exp
\left(g\int d\xi\ \bar\bpsi{(\xi)}
\, \gamma{(\xi)}\, \bpsi{(\xi)}\right)d\mu{(\gamma)}
$$

Performing the Fermionic integration
$$\eqalign{
\int e^{-\cVf}\, d\mu{(\bpsi,\bar\bpsi)}
\ &=\ \int \int
\exp
\left(g\int d\xi\ \bar\bpsi{(\xi)}
\, \gamma{(\xi)}\, \bpsi{(\xi)}\right)
d\mu{(\gamma_1,\gamma_2)}d\mu{(\bpsi,\bar\bpsi)}\cr
\ &=\ \int \det\big(\bbbone\!-\!g\, C\gamma\big)
d\mu{(\gamma)}\cr
}
$$
we obtain  (the exponential of) an effective interaction for the intermediate
boson field $\, \gamma\, $. Here, $\, \gamma\, $ is a multiplication
operator in position space acting on $\, \bbbr^{d+1}$-valued functions
and $\, C\, $
is the multiplication operator in momentum space given by
$$
C(p)
\ =\
-\rho(p){ip_0+e(\p)\sigma^3\over p_0^2+e(\p)^2}
$$
where $\rho(p)$ is the characteristic function of the set
$\ \set{p\in\bbbr^{d+1}}{p_0^2+e(\p)^2<1}.$ Thus $\rho(p)$
imposes an ultraviolet, but no infrared, cutoff on the Fermions.

To analyze the effective model  we introduce
(to be technically precise, already in the second regime) a
new scale that resolves the delta function covariance defining
the bosonic Gaussian measure $\, d\mu(\gamma)\, $.
This decomposition induces another renormalization group flow.

\vskip.15in
The determinant
$\
\det\big(\bbbone\!-\!g\, C\gamma\big)
\ $
is a complicated function of $\, \gamma\, $. To get some
feeling for it we do a mean field computation. We consider constant $\ga$'s and
introduce the periodized Fermionic covariance $P_j(\xi),$
$$\eqalign{
\ P_j(\xi)
\ &=\ \sum_c\ C(\xi\!-\!c)\cr
}
$$
The sum runs over the lattice $\ M^{-j}\bbbz^{d\!+\!1}$
so that $P_j$ is periodic on a large box $\Lambda$ of side $\, M^{-j}$. It
is not hard to show that: \hfil

{\it
\noindent
If $\, \gamma\, $ is constant on $\, \Lambda\, $, then
$$\eqalign{
\log\det(\bbbone-g \, P_j \gamma)
\ &=
\ |\Lambda|\, \sum_p\,{\tst {1\over |\Lambda|}}\, \log\left({\tst 1+
{g^2\ga^2\rho(p)\over p_0^2+e(\p)^2}}\right)\cr
}
$$
where the sum is over $\, p\, $ in $\, 2\pi M^j\bbbz^{d+1}\, $ and
where with abuse of notation
$\
\gamma^2\ =
\ (\gamma_{1}^2+\gamma_{2}^2)\bbbone_2
\ $
is identified with $\ (\gamma_{1}^2+\gamma_{2}^2)\ $.}
\remark{}{ Notice that
$\
\sum_p\, {1\over |\Lambda|}\ \rightarrow\ \int_{\bbbr^{d\!+\!1}}\dbar p
\ $
as $\ j\rightarrow -\infty\ $. On the other hand, the volume prefactor
$\ |\Lambda|=M^{-(d+1)j}\ $ tends to infinity as usual.}
\vskip.25truein
Formally,
$$\eqalign{
\det(\bbbone-g \, C \gamma)\, d\mu(\gamma)
\ &=\ e^{\log\det(\bbbone-g \, C \gamma)}
e^{-{1\over 2}\int d\xi\, \gamma B^{-1}\gamma}
\smprod_{\xi\in \bbbr^{d+1}}d\gamma{\sst (\xi)}\cr
\ &=\ e^{-{1\over 2}\big(\int d\xi\, \gamma(\xi)^2
-\log\det(\bbbone-g \, C \gamma)\big)}
\smprod_{\xi\in \bbbr^{d+1}}d\gamma{\sst (\xi)}\cr
}
$$
Thus, the full effective potential
in a box $\, \Lambda\, $ of side $\, M^{-j}\, $ evaluated at the constant
field configuration $\ \gamma\, $ is
$$
M^{-(d\!+\!1)j}\,
\Big({\textstyle{1\over 2}}\gamma^2
\, -\, \sum_p\, {1\over |\Lambda|}\log\left({\tst 1+g^2
{\rho(p)
\over p_0^2+e(\p)^2}
\, \gamma^2}\right)\Big)
$$
We want to show that its graph is a Bordeaux wine bottle (also referred to as
a Mexican hat) and determine its dimensions.

To do this it is convenient to replace the sum
$\ \sum_p\, {1\over |\Lambda|}\ $ by an
integral and study the mean field effective potential per unit volume
$$
{\cal E}(r)\ =\  {\textstyle{1\over 2}}r^2\, -\, \int\!\!\dbar p\
\log\left({\tst 1+g^2{\rho(p)\over p_0^2+e(\p)^2}\, r^2}\right)
$$
where
$\, r={\sst\sqrt{\gamma_1^2+\gamma_2^2}}\, $. In terms of the variable $s=r^2$
(but, by abuse of notation, retaining the name $\cE$)
$$\eqalign{
{\cal E}(s)\ &=\  {\textstyle{1\over 2}}s\, -\, \int\!\!\dbar p\
\log\left({\tst 1+g^2{\rho(p)\over p_0^2+e(\p)^2}\, s}\right)\cr
\frac{d\cE}{ds}(s)\ &=\  {\textstyle{1\over 2}}\, -\, \int\!\!\dbar p\
{ g^2\rho(p)\over p_0^2+e(\p)^2+g^2 s}\cr
\frac{d^2\cE}{ds^2}(s)\ &=\ \int\!\!\dbar p\
{g^4\rho(p)\over \left(p_0^2+e(\p)^2+ s\right)^2}\cr
}$$
Hence $\cE(s)$ is continuous on $[0,\infty)$, is zero at $s=0$ and grows
like $s/2$ at $s=\infty$. The first derivative diverges logarithmically to
$-\infty$ at $s=0$ and converges to $1/2$ at $s=\infty$. The second
derivative is always positive. Thus $\cE(s)$ has a unique critical point $s_*$
and this critical point is a global minimum.

Integrating over the angular variables, changing
variables to $\ \eta=e(\p)\ $ and then using polar coordinates to
replace $p_0$ and $\eta$
$$\eqalign{
{\cal E}(s)
\ &=\  {\textstyle{1\over 2}}s
-{\textstyle {|S^{d\!-\!1}|\over (2\pi)^{d\!-\!1}}}
\int \dbar p_0\dbar |\p|\, |\p|^{d\!-\!1}
\log\left({\tst 1+g^2{\rho(p)\over p_0^2+e(\p)^2}\, s}\right) \cr
&\ =\  {\textstyle{1\over 2}}s
-mk_F^{d-2}{\textstyle{|S^{d\!-\!1}|\over (2\pi)^{d\!-\!1}}}
\int \dbar p_0\dbar\eta
\ {\textstyle \left(1+{2\m\over k_F^2}\eta\right)^{{d-2\over 2}}}
\log\left({\tst 1+g^2{\rho(p)\over p_0^2+\eta^2}\, s}\right) \cr
&\ =\  {\textstyle{1\over 2}}s
-mk_F^{d-2}{\textstyle{|S^{d\!-\!1}|\over (2\pi)^d}}
\int_0^1 dR \ R\big(1+{\sst O(R^2)}\big)
\log\left(1+{g^2s\over R^2}\right) \cr
}$$
When $\ d=2\ $ the $\ O(R^2)\ $ term is absent. When $\ d>2\  $ we used
oddness to show that the $\ O(\eta)\ $ term vanishes.

For $\ d=2\ $ we can explicitly evaluate the integral to show that
the graph of the effective potential is a Bordeaux wine bottle whose
absolute minimum is at
$\ g|\ga|_*\,\approx\, \exp\left\{-{\pi\over mg^2}\right\}\ $ and has
depth approximately $\ \frac{m}{4\pi}(g|\ga|_*)^2\  $
and curvature at the minimum approximately $\ \frac{m}{\pi}g^2\ $. The picture
in dimensions $\,d>2\ $ is similar. Note that the depth,
$M^{-(d\!+\!1)j}g|\ga|_*$,  of the effective potential
$
M^{-(d\!+\!1)j}\,
\Big({\textstyle{1\over 2}}\gamma^2
\, -\, \sum_p\, {1\over |\Lambda|}\log\left({\tst 1+g^2
{\rho(p)
\over p_0^2+e(\p)^2}
\, \gamma^2}\right)\Big)
$
in the whole box is enormous due to the volume factor $M^{-(d\!+\!1)j}$. It is
deep enough to break the symmetry of the whole model.

Symmetry breaking forces the value of $\ga$ to be concentrated near some point
$\De/g=\sfrac{1}{g}(\De_1\si^1+\De_2\si^2)$ with $|\De|=\sqrt{\De_1^2+\De_2^2}
=g|\ga|_*$. The phase is determined by a boundary condition.
 Then it is natural
to shift $\ga$ by $\De/g$ and define the radial and tangential components
$$\eqalign{
\ga&\=\sfrac{1}{2\De^2}\tr (\ga\De)\De+\sfrac{1}{2\De^2}\tr(\ga\De^\#)\De^\#\cr
&\=\De/g+\ga_{\rm rad}\De/|\De|+\ga_{\rm tan}\De^\#/|\De|\cr
}$$
where $|\De|=\sqrt{\De^2}$. While $\ga_{\rm rad}$ and $\ga_{\rm tan}$
are globally defined they can only be interpreted as radial and
tangential components in a small neighbourhood of $\ga=\De/g$.
In the new variables the measure
$$\eqalign{
 e^{-\cVf}\, d\mu{\sst(\bpsi,\bar\bpsi)}
&\=\int e^{g\int d\xi\ \bar\bpsi{\scriptstyle(\xi)}
\, \gamma{\scriptstyle(\xi)}\, \bpsi{\scriptstyle(\xi)}}
d\mu{\sst(\gamma)}d\mu{\sst(\bpsi,\bar\bpsi)}\cr
&\=\const\int e^{g\int d\xi\ \bar\bpsi\gamma_s\bpsi
+\int d\xi\ \bar\bpsi\, \De\, \bpsi}
e^{-|\De|/g\int d\xi\,\ga_{\rm rad}}
d\mu{\sst(\gamma_s)}d\mu{\sst(\bpsi,\bar\bpsi)}\cr
&\=\const\int e^{g\int d\xi\ \bar\bpsi\gamma_s\bpsi}
e^{-|\De|/g\int d\xi\,\ga_{\rm rad}}
d\mu{\sst(\gamma_s)}d\mu_\De{\sst(\bpsi,\bar\bpsi)}\cr
}$$
where $\ga_s=\ga_{\rm rad}\De/|\De|+\ga_{\rm tan}\De^\#/|\De|$ is the shifted
field and $d\mu_\De$ is the Grassmann-Gaussian measure with covariance
$$
C_\De=\frac{\rho(k)}{ik_0-e(\k)\si^3-\De\rho(k)}
$$
Expanding the integral
$$
\int e^{g\int d\xi\ \bar\bpsi\gamma_s\bpsi}d\mu_\De{\sst(\bpsi,\bar\bpsi)}
=\det(\bbbone-gC_\De\ga_s)
$$
in powers of $\, g\, $ generates vertices whose naive power counting is
nonrenormalizable. We derive [FMRT4] quantitative Ward identities that force
the model to be superrenormalizable.

It is not straight forward
to rigorously exploit the structure of the effective potential and to
nonperturbatively implement the Ward identities. Roughly, at each
bosonic scale the Goldstone boson field is divided into a slowly varying
``background'' and a quickly varying fluctuation. Phase space is decomposed
into ``large'' and ``small'' field regions relative to the scale dependent
background fields. In small field regions the Goldstone boson field is
trapped ``near'' the bottom of the Bordeaux wine bottle and its gradient
is ``small''. Here, ``near'' and ``small'' quantitatively reflect the bosonic
scale. In large field regions, the Goldstone boson field is
``far'' from the bottom and its gradient
is ``large''. We expand, renormalize and apply Ward identities
in small field regions to extract the momentum space singularity symptomatic
of the long range nature of the Goldstone boson. We use probabilistic
extimates in large field regions to show that they are strongly suppressed.

\vskip.5truein
\noindent
\noindent{\subchfont References}
\vskip.1truein
\item{[A]} P.W. Anderson, Phys. Rev. Lett. {\bf 64}, 1839 (1990);
Phys. Rev. Lett. {\bf 65}, 2306 (1990)
\item{[deG]} P.G. deGennes, Superconductivity of Metals and Alloys,
Benjamin, New York, 1966
\item{[F]} J. Feldman, Introduction to Constructive Quantum Field Theory,
Proc. ICM Kyoto 1990, 1335-1341
\item{[FT1]} J. Feldman and E. Trubowitz, Renormalization in Classical
Mechanics and Many Body Quantum Field Theory, Journal d'Analyse
Math\'ematique {\bf 58} (1992) 213-247.
\item{[FT2]} J. Feldman and E. Trubowitz, Perturbation Theory for Many
Fermion Systems, Helvetica Physica Acta {\bf 63} (1990) 156-260.
\item{[FT3]} J. Feldman and E. Trubowitz, The Flow of an
Electron-Phonon System to the Superconducting
State, Helvetica Physica Acta, {\bf 64} (1991) 214-357.
\item{[FKLT]} J. Feldman, H. Kn\"orrer, D. Lehmann and E. Trubowitz,
Two Dimensional Fermi Liquids, to appear in {\it Constructive Results in
Field Theory, Statistical Mechanics and Solid State Physics},  edited by
V. Rivasseau
\item{[FMRT1]} J. Feldman, J. Magnen, V. Rivasseau and E. Trubowitz,
Constructive Many-Body Theory, in {\it The State of Matter}, M. Aizenmann
and H. Araki eds, Advanced Series in Mathematical Physics Vol.20,
World Scientific (1994).
\item{[FMRT2]} J. Feldman, J. Magnen, V. Rivasseau and E. Trubowitz,
Fermionic Many-Body Models, in {\it Mathematical Quantum Theory I: Field
Theory and Many-Body Theory}, J. Feldman, R. Froese and L. Rosen eds,
CRM Proceedings and Lecture Notes.
\item{[FMRT3]} J. Feldman, J. Magnen, V. Rivasseau and E. Trubowitz,
 An Infinite Volume Expansion for Many Fermion Green's Functions,
Helvetica Physica Acta, {\bf 65} (1992) 679-721.
\item{[FMRT4]} J. Feldman, J. Magnen, V. Rivasseau and E. Trubowitz,
Ward Identities and a Perturbative Analysis of a U(1) Goldstone Boson
in a Many Fermion System, Helvetica Physica Acta {\bf 66} (1993) 498-550.
\item{[FMRT5]} J. Feldman, J. Magnen, V. Rivasseau and E. Trubowitz,
An Intrinsic $1/N$ Expansion for Many-Fermion Systems,
Europhys. Lett., {\bf 24} (6), pp. 437-442 (1993)
\item{[FMRT6]} J. Feldman, J. Magnen, V. Rivasseau and E. Trubowitz,
Two-Dimensional Many-Fermion Systems as Vector Models,
Europhys. Lett., {\bf 24} (7), pp. 521-526 (1993)
\item{[H]} J. Woods Halley ed., Theories of High
Temperature Superconductivity,
Addison-Wesley, 1988

\end